\documentclass[sigplan,10pt,nonacm]{acmart}
\renewcommand\footnotetextcopyrightpermission[1]{}
\usepackage{multirow}
\usepackage{booktabs}
\usepackage{siunitx}
\usepackage{multicol}
\usepackage{graphicx}
\usepackage{caption}
\usepackage{subcaption}
\usepackage{tablefootnote}
\usepackage{enumitem}
\usepackage{amsmath}
\usepackage[capitalise]{cleveref}
\usepackage{xcolor}
\usepackage{tabularx}
\usepackage{amsfonts}
\usepackage{soul}
\usepackage[linesnumbered,ruled,vlined]{algorithm2e}
\usepackage{float}
\usepackage[indLines=true,italicComments=true,commentColor=gray]{algpseudocodex}
\usepackage{pgfplots}
\pgfplotsset{compat=1.18}
\usepackage{tcolorbox}
\usepackage{pifont}

\newcommand{\thiswork}{GF-DiT}

\begin{document}
\sloppy
\title{\thiswork{}: Scheduling Parallelism for Diffusion Transformer Serving}
% \hyphenpenalty=2500
% \widowpenalty=10000
% \clubpenalty=10000
\author{Xinwei Qiang}
\email{qiangxinwei@sjtu.edu.cn}
\affiliation{%
  \institution{Shanghai Jiao Tong University}
  \city{Shanghai}
  \country{China}
}

\author{Yifan Hu}
\email{hu_y_f@sjtu.edu.cn}
\affiliation{%
  \institution{Shanghai Jiao Tong University}
  \city{Shanghai}
  \country{China}
}

\author{Shixuan Sun}
\email{sunshixuan@sjtu.edu.cn}
\affiliation{%
  \institution{Shanghai Jiao Tong University}
  \city{Shanghai}
  \country{China}
}

\author{Jing Yang}
\email{jyang23@gzu.edu.cn}
\affiliation{%
  \institution{Guizhou University}
  \city{Guiyang}
  \country{China}
}

\author{Han Zhao}
\email{zhao-han@cs.sjtu.edu.cn}
\affiliation{%
  \institution{Shanghai Jiao Tong University}
  \city{Shanghai}
  \country{China}
}

\author{Chen Chen}
\email{chen-chen@sjtu.edu.cn}
\affiliation{%
  \institution{Shanghai Jiao Tong University}
  \city{Shanghai}
  \country{China}
}

\author{Yu Feng}
\email{y-feng@sjtu.edu.cn}
\affiliation{%
  \institution{Shanghai Jiao Tong University}
  \city{Shanghai}
  \country{China}
}

\author{Jingwen Leng}
\email{leng-jw@sjtu.edu.cn}
\affiliation{%
  \institution{Shanghai Jiao Tong University}
  \city{Shanghai}
  \country{China}
}

\author{Minyi Guo}
\email{guo-my@sjtu.edu.cn}
\affiliation{%
  \institution{Shanghai Jiao Tong University}
  \city{Shanghai}
  \country{China}
}

\begin{abstract}
Diffusion Transformers (DiTs) have become the dominant architecture for image and video generation, creating growing demand for efficient DiT serving. Existing systems assign each request a fixed parallel configuration throughout its lifetime. However, DiT workloads exhibit substantial heterogeneity across requests, execution stages, and system conditions, making static parallelism inefficient and often leading to poor GPU utilization and degraded service quality.

This paper argues that DiT serving should treat GPU parallelism as a first-class schedulable resource. We present \textbf{\thiswork{}}, a policy-programmable runtime for elastic DiT serving that dynamically adapts the parallelism of running requests according to workload demands and service objectives. \thiswork{} introduces an asynchronous execution abstraction that decomposes requests into independently schedulable trajectory tasks and enables online GPU reallocation. To make elastic parallelism practical, \thiswork{} further proposes group-free collectives, a lightweight communication abstraction that supports low-overhead online formation and reconfiguration of arbitrary execution groups.

We implement \thiswork{} in vLLM-Omni and evaluate it on representative image and video diffusion workloads. Compared with fixed-pipeline execution with static parallelism, \thiswork{} improves throughput by up to \textbf{6.01}$\times$, reduces mean latency by up to \textbf{95}\%, lowers SLO violation rates by up to \textbf{90}\%, and reduces communication-group setup overhead from \textbf{778} ms to approximately \textbf{60} $\mu$s.

Our code is available at \url{https://github.com/SJTU-Liquid/GF-DiT}.
\end{abstract}

\maketitle

\section{Introduction}

DiTs~\cite{DiT} have emerged as the dominant
architecture for modern generative AI workloads, enabling high-quality
text-to-image, text-to-video, and multimodal content
generation~\cite{wan,hunyuan,longcatvideo,vdt,gupta2023photorealisticvideogenerationdiffusion,ma2025lattelatentdiffusiontransformer,videoworldsimulators2024,zheng2024opensorademocratizingefficientvideo,yang2025cogvideoxtexttovideodiffusionmodels,polyak2025moviegencastmedia,zhou2024allegroopenblackbox,zheng2026opensora20trainingcommerciallevel,hacohen2024ltxvideorealtimevideolatent,gao2025seedance10exploringboundaries}.
As demand for T2I and T2V services continues to grow, efficiently serving DiT
models has become a critical systems challenge.

Unlike large language models (LLMs), which process requests through a prefill stage followed by autoregressive decoding, DiT inference consists of three stages: an \emph{encoder} that transforms prompts into embeddings, a compute-intensive \emph{denoising} stage that iteratively refines latent states through tens of diffusion steps, and a \emph{decoder} that generates the final image or video. Among these stages, denoising dominates execution time and resource consumption.

\begin{figure}[t]
\centering
\includegraphics[width=\linewidth]{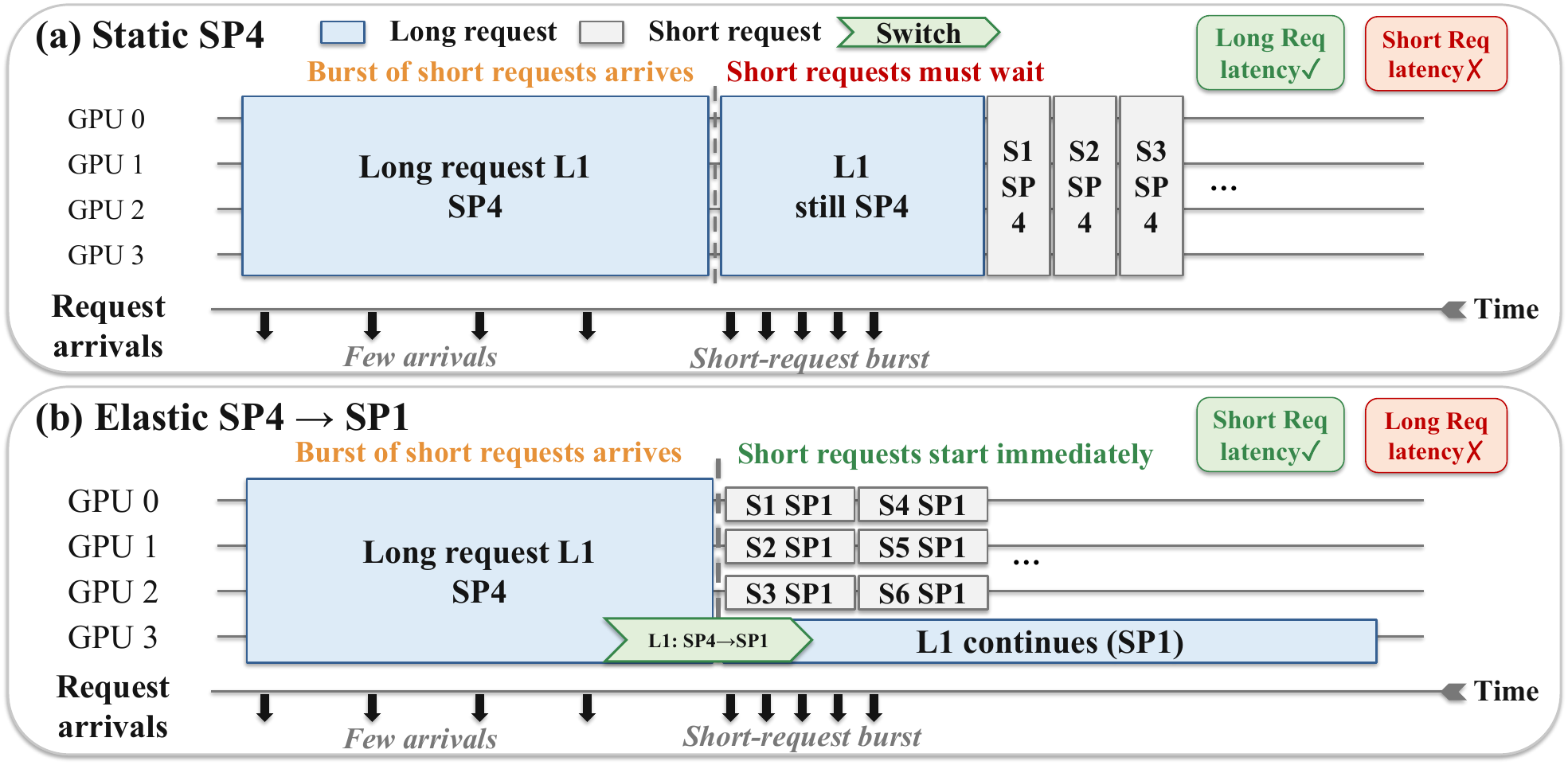}
\caption{Elastic parallelism exposes policy tradeoffs. Static SP4 preserves the long request’s latency but makes short requests wait; shrinking it to SP1 admits short requests sooner at the cost of delaying the long request.}
\label{fig:intro-elastic-parallelism}

\end{figure}

To serve DiT workloads, recent systems such as vLLM-Omni~\cite{vllm-omni} and
SGLang Diffusion~\cite{sglang-diffusion} adopt a fixed multi-GPU execution
strategy. As illustrated in \cref{fig:intro-elastic-parallelism}, each request
is assigned a predetermined parallel configuration, and the encoder, denoising,
and decoder stages are executed atomically under the same resource allocation.
However, DiT workloads exhibit substantial heterogeneity. Across requests,
output resolution, video duration, and the number of denoising steps vary
significantly, resulting in orders-of-magnitude differences in computational
demand. Within a request, different stages expose fundamentally different
degrees of parallelism: encoder and decoder stages are lightweight, while
denoising can effectively utilize substantially more GPUs.

Consequently, fixed-pipeline execution with static parallelism leads to two fundamental limitations. First, long-running requests introduce head-of-line (HoL) blocking, delaying shorter requests and degrading service quality. Second, static parallelism often leads to poor resource efficiency. Lightweight stages and small requests cannot effectively utilize many GPUs, whereas insufficient parallelism prolongs large denoising workloads. As a result, existing systems struggle to simultaneously achieve high SLO attainment and high GPU utilization.

At the root of these limitations is the static nature of parallelism in existing DiT serving systems. Once a request is assigned a parallel configuration, its GPU allocation remains unchanged throughout execution. Consequently, the system cannot react to workload dynamics such as request arrivals, changing queue conditions, or evolving service objectives. A request that is appropriately provisioned when it begins execution may later become either over-provisioned or under-provisioned as system conditions evolve.

These observations motivate a different perspective: \textbf{instead of treating parallelism as a fixed execution configuration, a DiT serving system should treat parallelism as a first-class schedulable resource.} Similar to how operating systems dynamically allocate CPU time among competing processes, a DiT serving system should dynamically allocate GPU parallelism among competing requests according to workload demands and service objectives. Under this abstraction, GPU resources can continuously flow to where they provide the greatest benefit. In other words, existing systems schedule requests, whereas a DiT serving runtime should additionally schedule parallelism.

However, realizing this vision efficiently is challenging. Dynamic parallelism adaptation requires the runtime to repeatedly reconfigure resource allocations among concurrently executing requests while minimizing disruption to ongoing computation. Moreover, different deployments often optimize for different objectives, including latency, throughput, fairness, and cost efficiency. Therefore, the challenge is not only to enable efficient runtime reconfiguration, but also to provide a programmable substrate upon which diverse scheduling policies can be built.

\textbf{Our Work.}
To address these challenges, we present \textbf{\thiswork{}}, a policy-programmable runtime for elastic DiT serving. \thiswork{} is enabled by two unique properties of DiT workloads. First, DiT execution maintains a \emph{lightweight execution state}: the data exchanged between stages and denoising steps consists primarily of latent states and embeddings whose sizes are typically only several MBs to tens of MBs. Second, DiT requests exhibit a \emph{predictable execution structure}: the denoising trajectory is largely determined before execution begins, enabling accurate estimation of execution costs and future resource demands. Together, these properties make DiT serving uniquely amenable to online parallelism adaptation, as requests can be reconfigured frequently while their future execution costs remain predictable.

Building upon these insights, \thiswork{} consists of three key components. First, we design an \emph{asynchronous execution abstraction and runtime model} that decomposes a DiT request into independently schedulable trajectory tasks, including encoding, decoding, and individual denoising steps. Each task boundary serves as an explicit rescheduling point, allowing the runtime to continuously adapt GPU allocations throughout request execution. By exposing rescheduling opportunities across the entire request lifetime, \thiswork{} elevates parallelism from a static deployment decision to a runtime-managed resource.

Second, elastic parallelism fundamentally requires execution groups to be created and reconfigured online. To enable this capability, we introduce \emph{group-free collectives}, a lightweight communication abstraction that eliminates expensive communicator construction and supports arbitrary GPU-group formation through lightweight logical descriptors. Group-free collectives make dynamic parallelism practical by reducing runtime reconfiguration overheads to a negligible level.

Finally, we develop a set of runtime optimizations that improve programmability
and deployment efficiency, including layout-aware artifact migration and
simulation-driven policy optimization. Combined with \thiswork{}'s programmable
policy interface, these techniques allow users to rapidly evaluate and deploy
customized scheduling strategies while reusing the same runtime substrate.

We implement \thiswork{} in vLLM-Omni and evaluate it on both image and video diffusion workloads. The same runtime substrate supports throughput-oriented, latency-oriented, SLO-aware, and custom scheduling policies through a unified policy interface. Compared with fixed-pipeline execution with static parallelism, \thiswork{} improves throughput by up to \textbf{6.01}$\times$, reduces mean latency by up to \textbf{95}\%, lowers SLO violation rates by up to \textbf{90}\%, and reduces communication-group setup overhead from up to \textbf{778} ms to approximately \textbf{60} $\mu$s. In summary, this paper makes the following contributions:

\begin{itemize}

\item We introduce \emph{parallelism as a first-class schedulable resource}, a new abstraction for adaptive DiT serving.

\item We design \textbf{\thiswork{}}, a policy-programmable runtime for elastic DiT serving that enables online GPU reallocation through asynchronous execution and reschedulable trajectory tasks.

\item We propose \emph{group-free collectives}, which make dynamic execution-group formation practical by eliminating expensive communicator construction.

\item We implement \thiswork{} and demonstrate significant performance improvements
over existing fixed-pipeline execution with static parallelism.

\end{itemize}

\section{Background and Motivation}

This section presents the key observations that motivate \thiswork{}. We first show
that DiT workloads naturally expose rescheduling boundaries throughout
execution. We then show that their execution structure is largely predictable
before execution begins. Finally, we demonstrate that no single parallel
configuration is optimal across stages, requests, and system conditions,
motivating our central thesis: GPU parallelism should be treated as a
first-class schedulable resource.

\subsection{Rescheduling Boundaries}
\label{sec:execution-units}

LLM and diffusion serving both execute requests iteratively, but they expose fundamentally different opportunities for runtime scheduling. In autoregressive LLM serving, each generated token produces a valid continuation state, while chunked prefill incrementally constructs reusable KV caches~\cite{sarathi}. These token-sequence boundaries naturally provide opportunities for batching, preemption, and request scheduling.

DiT execution follows a different structure. During denoising, the model repeatedly refines a latent state along a \emph{diffusion trajectory}, i.e., the sequence of latent states traversed from an initial noisy latent to the final clean latent. Within a denoising step, the model performs bidirectional computation over the entire latent sequence, and an arbitrary latent-token prefix does not represent semantically complete progress along the diffusion trajectory. Although a DiT implementation may internally employ sequence parallelism or model-specific partitioning, such boundaries are implementation artifacts rather than portable scheduling points that can be safely exposed to a serving runtime.

Instead, DiT requests expose natural scheduling opportunities at stage and denoising-step boundaries. A request first executes conditioning encoders (e.g., text or image encoders) to produce embeddings and initialize latent states. The denoising stage then repeatedly applies the DiT model across a sequence of timesteps, progressively advancing the diffusion trajectory. Finally, a VAE~\cite{vae} decoder converts the denoised latent into the final image or video.

\begin{figure}
\centering
\includegraphics[width=\linewidth]{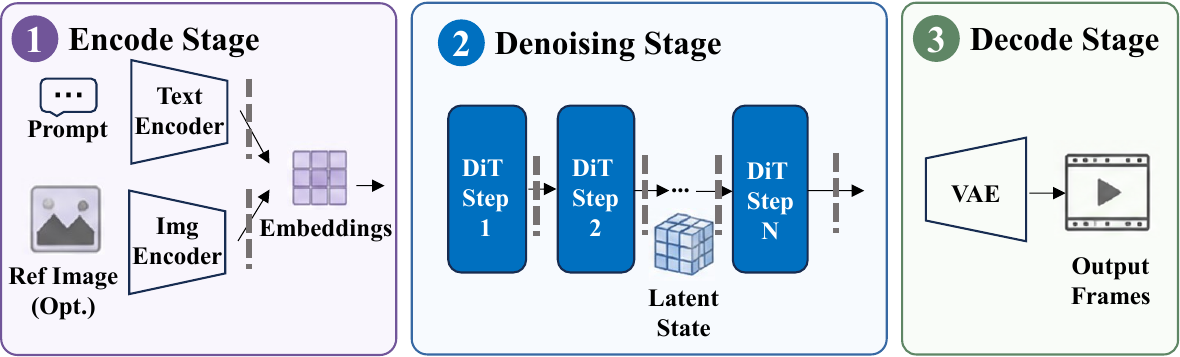}
\caption{Structure of a diffusion serving request. Encoding produces conditioning embeddings, denoising iteratively advances the diffusion trajectory, and VAE decoding generates the final output. Trajectory task boundaries provide semantically valid rescheduling points for runtime adaptation.}
\label{fig:diffusion-request-structure}
\end{figure}

As illustrated in \cref{fig:diffusion-request-structure}, completing an encoder
stage, a decoder stage, or a denoising step produces a semantically complete
execution state that can be safely transferred, resumed, or executed under a
different resource configuration. Each such point is a \emph{rescheduling
boundary} at which the runtime may reconsider task placement and GPU allocation.
\thiswork{} represents each stage or denoising step between these boundaries as an
independently schedulable \emph{trajectory task}. By exposing these boundaries
throughout request execution, \thiswork{} creates opportunities for online
parallelism adaptation while preserving request semantics.

\subsection{Predictable Execution Structure}
\label{sec:predictable-execution-structure}

A key property of DiT workloads is that much of a request’s execution structure is known before execution begins. Request parameters or service defaults specify the output resolution, video duration, and denoising-step count at admission time. For a fixed model and pipeline, these parameters determine the latent sequence length and the overall diffusion trajectory.

Once a request is decomposed into trajectory tasks, its execution graph becomes largely deterministic. The number of tasks, their dependency structure, and the remaining execution path are all known a priori. Moreover, each trajectory task corresponds to a known model stage or denoising step operating on a known input shape. Its execution cost under a candidate parallel configuration can therefore be estimated from profiling data rather than discovered only after execution completes.

This predictability fundamentally distinguishes DiT serving from many dynamic serving workloads. Because future execution costs remain largely predictable, the runtime can reason about alternative resource allocations before they are applied. Combined with the rescheduling opportunities provided by trajectory task boundaries, this property makes online parallelism adaptation both practical and effective.

\begin{figure*}
\centering
\includegraphics[width=\textwidth]{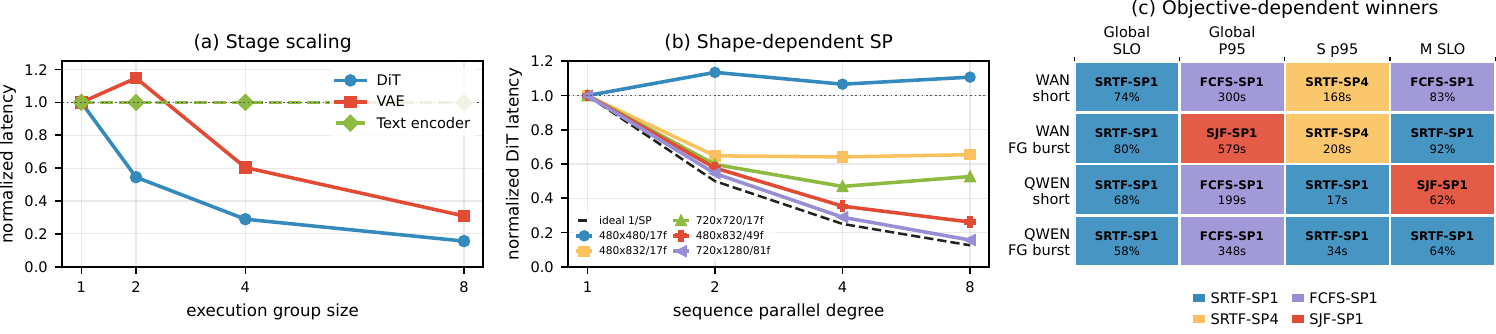}
\caption{Motivating measurements for elastic DiT serving. (a) Different stages exhibit distinct scaling behavior and resource preferences. (b) The performance benefit of parallelism depends on request shape. (c) Different workload conditions favor different parallelism choices, indicating that no single parallel configuration is universally optimal.}
\label{fig:motivation}
\end{figure*}

\subsection{Why Parallelism Should Be Schedulable}
\label{sec:motivation}

We next use microbenchmarks and trace-driven replay to show why fixed parallelism is fundamentally insufficient for DiT serving. Our results demonstrate that the optimal degree of parallelism depends on the execution stage, request characteristics, and system conditions.

\textbf{Stage-level heterogeneity.}
\cref{fig:motivation}(a) reports representative stage latencies under different execution-group sizes. Different stages exhibit fundamentally different scaling behavior. Denoising tasks benefit from larger sequence-parallel groups because they dominate computation and operate on long latent sequences. In contrast, text encoding remains effectively single-rank, while VAE decoding exhibits a distinct scaling profile. Consequently, a single request-wide parallel configuration either wastes resources on lightweight stages or under-provisions compute-intensive denoising tasks.

\textbf{Shape-dependent parallelism.}
\cref{fig:motivation}(b) reports the latency of a denoising task, normalized to single-GPU execution, across request shapes and sequence-parallel degrees. Larger image and video requests expose more computation per denoising step and therefore benefit more from additional GPUs. Smaller requests, however, often fail to amortize communication and synchronization overheads, resulting in limited gains from larger execution groups. Even for the same model and execution stage, the optimal parallel configuration depends strongly on request shape.

\textbf{System-dependent parallelism.}
\cref{fig:motivation}(c) evaluates different serving configurations using trace-driven replay. Under lightly loaded conditions, allocating more GPUs to individual requests may minimize latency. Under heavier workloads, however, reducing per-request parallelism often improves overall throughput and SLO attainment by increasing concurrency. As workload composition, request arrivals, and service objectives evolve over time, the preferred parallel configuration changes accordingly.

Together, these observations show that no single parallel configuration is optimal throughout a request’s lifetime. Different execution stages, request characteristics, and system states prefer different degrees of parallelism. This motivates the central design principle of \thiswork{}: rather than treating parallelism as a fixed execution configuration chosen at admission time, a DiT serving system should treat \emph{GPU parallelism as a first-class schedulable resource}. Trajectory task boundaries create opportunities for adaptation, predictable execution structure makes adaptation feasible, and workload heterogeneity makes adaptation necessary. Enabling such adaptation efficiently motivates the asynchronous execution abstraction, group-free collectives, and policy-programmable runtime presented in the remainder of the paper.

\section{\thiswork{} Abstraction and Overview}
\label{sec:abstraction}

\thiswork{} turns parallelism from a static deployment choice into a
schedulable resource. It exposes runtime-visible trajectory tasks along each
diffusion trajectory, lets a policy bind each ready task to an execution layout, and
realizes the resulting decisions through an asynchronous runtime. This section
defines the abstraction and shows how its components fit together;
\cref{sec:gfc,sec:implementation} describe the communication mechanism and
runtime implementation.

\subsection{Reschedulable Trajectory Tasks}

\thiswork{} represents each diffusion request as a placement-agnostic
\emph{trajectory task graph}. Its nodes are independently schedulable trajectory
tasks, and its edges are artifact dependencies. A trajectory task may represent
a model stage, such as encoding, latent preparation, or decoding, or a
single denoising step. Completing a task produces a well-defined model state,
so the runtime may reconsider task order, placement, and parallelism at the next
boundary without exposing partially updated model internals.

Trajectory tasks communicate through \emph{logical artifacts}. An artifact may
represent conditioning features, latent state, scheduler metadata, decoded
outputs, or other model-specific state needed by downstream tasks. At the
abstraction level, an artifact records a dependency and semantic role but not a
physical layout. The same artifact may later be materialized as replicated
tensors, sequence-parallel shards, or model-specific metadata according to the
layouts selected for its producer and consumer.

This graph exposes two properties of diffusion execution to the runtime.
First, task boundaries are explicit rescheduling points at which GPU allocation
can change. Second, the graph and request shape are largely known at admission,
allowing the runtime to estimate remaining work and compare candidate parallel
configurations before a task executes.

\subsection{Execution Layouts and Policy Interface}

A \thiswork{} policy schedules a ready trajectory task by assigning it an
\emph{execution layout}. An execution layout consists of an ordered logical
execution group and a parallel specification describing how the task uses that
group. For example, a DiT task may use a multi-rank sequence-parallel layout,
whereas a lightweight encoder task may use a single rank.

At each scheduling point, the policy observes ready tasks, request metadata,
resource availability, optional deadlines or priorities, and cost estimates for
candidate layouts. It returns dispatch decisions of the form
\[
(task, execution layout).
\]
This interface lets policies jointly control task order, placement, and
parallelism. A policy may prioritize short requests, allocate additional ranks
to an urgent request, or shrink a running request at its next trajectory
boundary to admit new work. The policy operates only on logical tasks, artifacts, and execution groups.
It does not construct communicators, invoke model stages, or plan tensor
transfers. The runtime validates each decision and handles these execution
mechanics. In particular, group-free collectives make policy-selected logical
groups executable without serving-path communicator construction, while
layout-aware artifact migration reconstructs artifacts when consecutive tasks use
different layouts.

\subsection{System Overview}

\begin{figure}
\centering
\includegraphics[width=\linewidth]{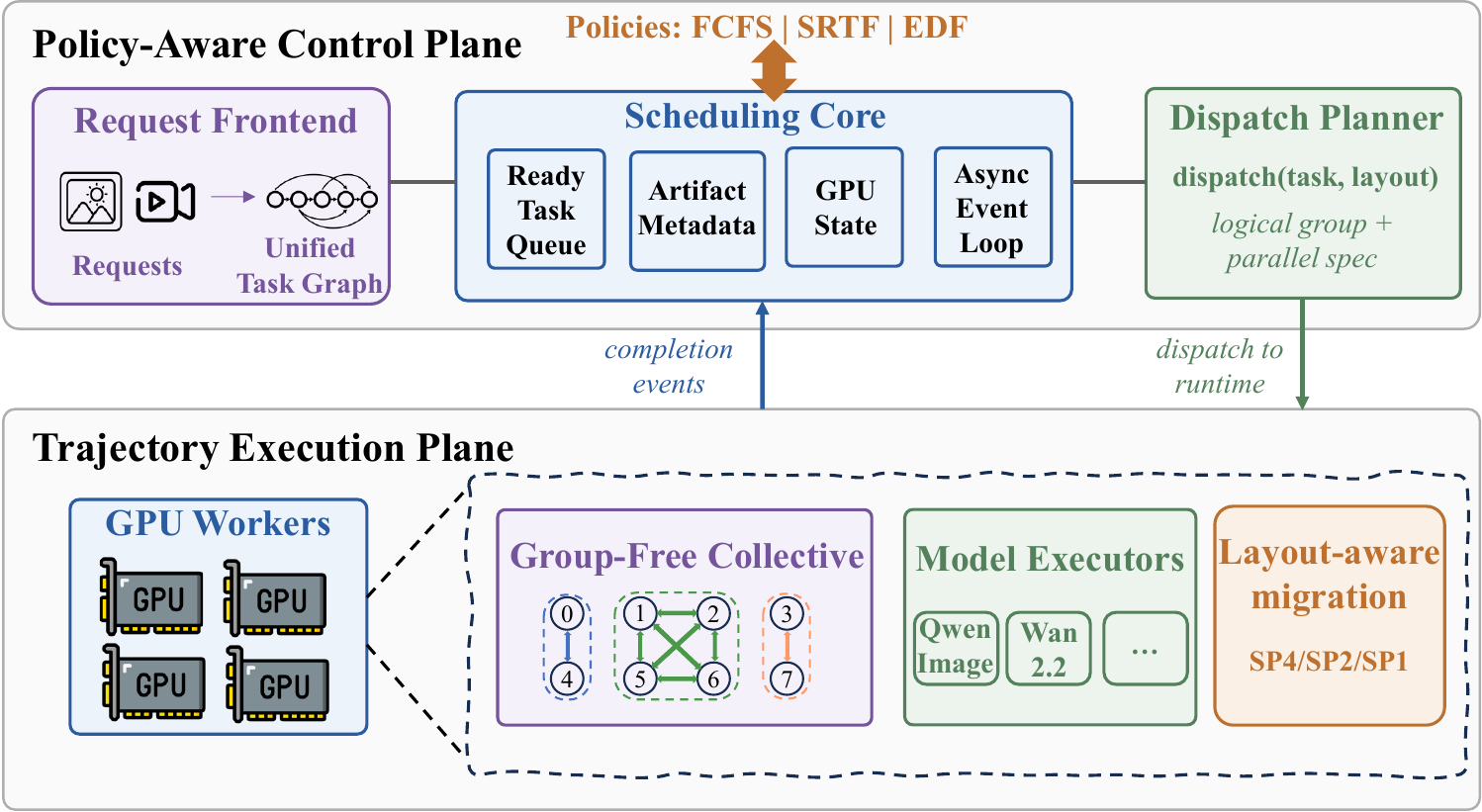}
\caption{\thiswork{} system overview. Incoming requests are converted into
trajectory task graphs, scheduled by the control plane through a programmable
policy interface, and dispatched to workers under dynamic execution layouts.
Group-free collectives and layout-aware artifact migration make these dynamic layout
choices executable at runtime.}
\label{fig:overview}
\end{figure}

\cref{fig:overview} shows how the abstraction maps to the system. A
model-specific adapter converts each incoming request into trajectory tasks and
logical artifacts. The control plane maintains graph dependencies, artifact
metadata, resource state, and policy-visible request information. When tasks
become ready or resources are released, it invokes the policy and dispatches
selected tasks under their assigned execution layouts. The execution plane consists of GPU workers, model executors, and communication
runtimes. Workers execute tasks asynchronously and report progress to the
control plane, allowing scheduling and CPU-side preparation to overlap with GPU
execution. Group-free collectives support dynamic execution groups, and
layout-aware artifact migration reconstructs intermediate artifacts when a
policy changes placement or parallelism across task boundaries.

The same control-plane abstraction also supports offline policy exploration.
The online runtime completes dispatched tasks through GPU workers; the simulator
instead produces completion events from profiled task costs. Because both
backends preserve the same task graph, resource state, and policy interface,
policies evaluated in simulation can be deployed without rewriting their
decision logic.

\section{Group-Free Collectives}
\label{sec:gfc}

The abstraction in \cref{sec:abstraction} lets policies assign
trajectory tasks to dynamic execution layouts. DiT tasks often need subgroup
collectives within those layouts, such as all-gather or all-to-all operations
used by sequence-parallel execution~\cite{vllm-omni}. \thiswork{} executes these
operations with group-free collectives (GFC): the runtime pays one world-level
communication setup cost at initialization and represents each dynamic subgroup
as lightweight metadata.

\begin{table}[t]
  \centering
  \caption{NCCL subgroup setup costs on an 8-GPU server.}
  \label{tab:nccl-subgroup-cost}
  \scriptsize
  \setlength{\tabcolsep}{3.2pt}
  \begin{tabular}{r@{\quad}rrrr}
    \toprule
    Group & \texttt{new\_group} & First coll. & Warm coll. & Memory \\
    size & (ms) & (ms) & (ms) & (MB/GPU) \\
    \midrule
    2 & 0.77 & 217.3 & 0.19 & 484 \\
    4 & 0.46 & 532.0 & 0.56 & 535 \\
    6 & 0.57 & 663.5 & 0.72 & 535 \\
    8 & 0.57 & 777.7 & 0.89 & 535 \\
    \bottomrule
  \end{tabular}
\end{table}

\subsection{Limits of Process-Group-Based Collectives}

Conventional collective libraries expose subgroup communication through
communicators or process groups. A process group fixes the participating ranks
and provides the state needed by collective operations issued within that
group. This model works well for static model-parallel deployments, where the
same tensor-parallel or sequence-parallel groups are reused for the lifetime of
the server.

Policy-programmable diffusion serving changes this assumption. A throughput
policy may run many requests on small groups to increase concurrency, while a
deadline-aware policy may temporarily allocate a larger group to an urgent
request. A single request may also use different groups for different points on
its trajectory: encoder stages may run on one rank, early DiT tasks may run on
a larger sequence-parallel group, and later tasks may shrink or move as load
changes. These decisions create subgroup rank sets that are chosen online
rather than fixed at server startup.

Creating a process group for each selected rank set is a poor fit for the
serving path. Construction involves host-side coordination and communicator
initialization, so it adds latency exactly when the scheduler is making a
fine-grained placement decision. Pre-creating candidate groups removes
serving-path setup, but the number of subgroups grows exponentially with rank
count; even a restricted set consumes memory and limits the policy to choices
known ahead of time. For example, pre-creating every size-2 and size-4 group
already requires $\binom{8}{2}+\binom{8}{4}=98$ process groups on an 8-GPU
server. On an NVL72 domain, the same restricted set grows to
$\binom{72}{2}+\binom{72}{4}=1{,}031{,}346$ groups. Given the nontrivial
per-group memory footprint measured below, exhaustive pre-creation is
impractical even before considering other group sizes.

\cref{tab:nccl-subgroup-cost} quantifies this cost on an 8-GPU server. The
explicit \texttt{new\_group} call returns in sub-millisecond time, but the first
collective on the subgroup triggers hundreds of milliseconds of cold-path
initialization and reserves roughly 0.5~GB of GPU memory. Warm collectives are
fast; the problem is the one-time setup and memory footprint that dynamic
policies would pay for each newly selected rank set.

\thiswork{} therefore separates execution groups from communicators. A policy
chooses a logical execution group as part of a task's layout, while the
communication runtime executes collectives for that group using GFC descriptors
and world-level communication state.

\subsection{Scope and Ordering Assumptions}
\label{sec:gfc-scope}
GFC targets a controlled serving setting rather than arbitrary user-issued
collective communication. The control plane decides which trajectory tasks run,
which ranks participate in each layout, and which subgroup collectives those
tasks issue. Executors invoke collectives through the runtime interface instead
of directly constructing communicators, allowing \thiswork{} to support dynamic
subgroups without exposing a general distributed communicator API.

The main correctness requirement is ordering. A rank may participate in many
logical groups over time, and these groups may overlap. For example, ranks
0 and 1 may first communicate as part of group \(\{0,1,2,3\}\) and later as
part of group \(\{0,1\}\). If rank 0 enters these shared collective instances
in a different order from rank 1, then a synchronization signal from one
instance may be mistaken for another. This can lead to incorrect data movement
or deadlock~\cite{nccl}.

\thiswork{} therefore assumes \emph{pairwise-consistent ordering}: for any pair
of ranks, the collective instances in which both ranks participate are issued
in the same order at both ranks. This assumption is weaker than requiring a
single global order over all collectives. Two disjoint groups may proceed
independently, and groups that overlap only partially need only agree on the
relative order observed by each shared rank pair. Under this runtime model, \thiswork{} enforces pairwise-consistent ordering
through centralized scheduling and ordered submission. The scheduler is the only
component that creates dynamic execution layouts, and each worker submits
collectives for assigned tasks to a local ordered communication stream. Thus,
any two ranks that appear together in multiple groups observe the same order of
their shared collective instances.

\begin{figure*}[t]
\centering
\includegraphics[width=\textwidth]{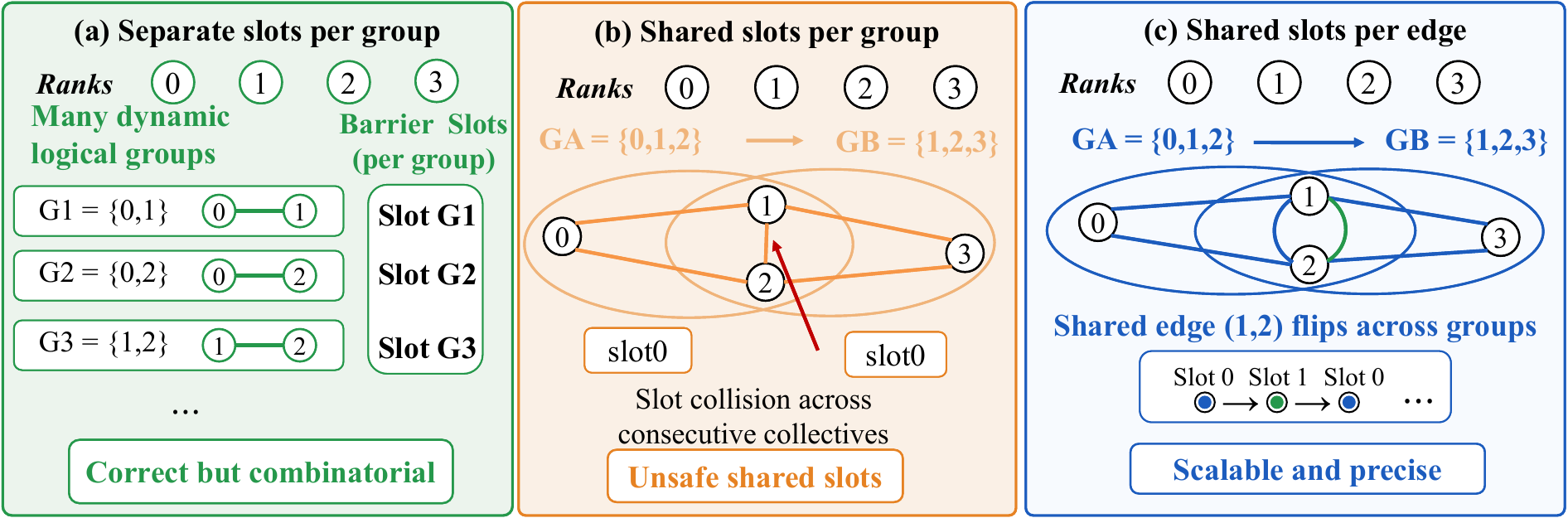}
\caption{Signal-state designs for dynamic overlapping groups. (a) Separate
slots for every logical group avoid collisions but require combinatorial
per-group state. (b) Sharing slots across groups allows consecutive overlapping
collectives to collide. (c) \thiswork{} assigns double-buffered phase state to
each rank edge, so a shared edge flips slots consistently across groups without
allocating per-group signal state.}
\label{fig:gfc-edge-state}
\end{figure*}

\subsection{Symmetric Buffers and Logical Groups}

Group-free collectives separate communication state from subgroup membership.
At initialization, \thiswork{} performs a single world-level setup in which all
ranks allocate and register symmetric communication buffers~\cite{pytorch-symmetric-memory}. These buffers
provide a common addressable substrate for data movement and synchronization
across ranks. This setup is paid once when the server starts, rather than each
time a policy creates a new execution group.

At runtime, a subgroup is represented by a logical group descriptor containing
the ordered participating ranks, a runtime group identifier, and the local
rank's position in that order. Creating a logical group is therefore a metadata
operation: the runtime constructs no communicator, allocates no heavy per-group
communication state, and does not require non-members to join. A trajectory task
can use a logical group once the participating ranks receive the same descriptor
from the control plane. In our implementation, registration takes approximately
\textbf{60}~$\mu$s: CPU-side descriptor bookkeeping plus one host-to-device copy
of the participating rank IDs, with no communicator initialization or subgroup
warmup.

The descriptor gives each rank enough information to interpret collective
semantics inside the group. For example, it defines the local rank index used by
a sequence-parallel DiT executor and maps group-local peers to global ranks for
communication. The symmetric buffers provide the storage and addressing needed
to exchange data with those peers. Together, the descriptor and the world-level
buffers replace the role that a per-subgroup process group would normally play
in dynamic execution.

\thiswork{} keeps this interface separate from model tensors. Executors operate
on ordinary input and output tensors; when a collective is issued, the
communication runtime stages data into a symmetric buffer region, performs
subgroup data movement, and copies results back if needed. This staging hides
GFC from model code while allowing the runtime to choose different backends for
the same logical group abstraction.

\subsection{Edge-Based Agreement for Overlapping Groups}

Logical group descriptors and symmetric buffers make subgroup communication
addressable, but they do not establish agreement on collective instances. Since
logical groups can overlap, the same pair of ranks may communicate together in
different groups over time. The runtime must ensure that a signal written by one
rank for one collective is interpreted by its peer as belonging to that same
collective.

\thiswork{} identifies each collective instance with a synchronization token
derived from the runtime session, the logical group identity, and a per-group
logical epoch. The epoch advances whenever the group issues a new collective.
The token therefore tells a peer which collective instance a signal belongs to,
even if the same signal slot is reused later.

The remaining problem is deciding when a signal slot can be reused. A single
slot per sender-receiver pair is unsafe: a rank could publish the next token
before its peer has observed the previous one, overwriting the only evidence of
arrival. Double buffering fixes this local lifetime problem by alternating
between two slots, so instance \(N+1\) does not overwrite the token for instance
\(N\). However, the phase bit that selects the slot must advance in the same way
at both endpoints for exactly the collective instances they share.

Two group-level slot designs are unsatisfactory. Giving every possible logical
group its own barrier slots is correct, but the number of possible groups is
combinatorial. Reusing one shared slot pool across logical groups is compact but
unsafe: consecutive overlapping groups can make their shared rank pair publish
to the same slot before the earlier token has been consumed, as in
\cref{fig:gfc-edge-state}(b). The slot sequence must therefore be owned by
the entity that actually observes reuse, namely the rank pair.

\thiswork{} attaches the phase bit to each ordered rank edge. A group barrier
requires each rank to agree with every other rank in the logical group and
completes only after all pairwise edges observe the expected token.
\cref{fig:gfc-edge-state} contrasts this design with group-level
alternatives. Because slot ownership follows rank pairs, unrelated groups cannot
overwrite one another, and overlapping groups interact only through the edges
whose endpoints actually need to agree.

\begin{algorithm}[t]
\caption{Per-edge flip agreement for a GFC collective}
\label{alg:gfc-agreement}
\KwIn{Logical group $G$, local rank $r \in G$, collective instance $c$}
\textbf{State:} each local edge $(r,p)$ keeps a one-bit phase
$\phi_{r,p}$ and two signal slots\;
\textbf{Invariant:} for every rank pair, shared collective instances are
submitted in the same order at both endpoints\;

$\tau \gets \textsc{Token}(G,c)$\;

\ForEach{$p \in G \setminus \{r\}$}{
    $e \gets (r,p)$\;
    $s[p] \gets \phi_{r,p}$\;
    $\phi_{r,p} \gets 1 - \phi_{r,p}$\tcp*{Flip phase}
    $\textsc{Publish}(e,s[p],\tau)$\;
}

\ForEach{$p \in G \setminus \{r\}$}{
    $e' \gets (p,r)$\;
    wait until $\textsc{Observed}(e',s[p],\tau)$\;
}

\end{algorithm}

Algorithm~\ref{alg:gfc-agreement} intentionally hides the signal layout. In
the implementation, each ordered rank pair owns two signal slots. A rank
publishes to the peer's incoming slot and waits on the reciprocal slot. The
slot is selected by the parity of a local per-edge sequence counter. Because
the control plane submits pair-sharing collectives in the same order at both
endpoints, the two ranks select matching slots for the same collective
instance without exchanging slot state.

The protocol requires no separate acknowledgement phase. Because each rank
submits edge instances on one ordered communication stream, a slot used by
edge instance \(N\) is not reused until instance \(N+2\). Before a rank can
publish \(N+2\), its local \(N+1\) instance must have returned; that return
implies that the peer published \(N+1\), which occurs only after the peer
consumed the token for \(N\). The double buffer is therefore sufficient to
prevent a token from being overwritten before its peer observes it. Publish
and observe use system-scope release and acquire operations, respectively, so
data movement begins only after all pairwise arrivals are visible.

This design relies on the pairwise-consistent ordering assumption from
\cref{sec:gfc-scope}. Under that assumption, the next signal on an edge
corresponds to the same collective instance at both endpoints, while the token
detects stale or mismatched observations. \thiswork{} satisfies the assumption
because the centralized control plane creates dynamic groups and workers submit
collectives through ordered per-rank communication streams.

\subsection{Backend-Aware Collective Execution}

After a logical group has been registered and its collective instance reaches
agreement, \thiswork{} executes data movement with a backend chosen for the
message size and communication pattern. The GFC interface is backend
independent: executors issue subgroup collectives over ordinary tensors, while
the communication runtime maps each operation to CUDA-kernel communication,
copy-engine transfers, or TMA-style transfers when available. This separation
keeps model code independent of the communication mechanism.

A practical challenge is that executor tensors are not required to reside in
symmetric communication buffers. Requiring all model intermediates to be
allocated from a special symmetric-memory pool would make integration intrusive
and would complicate existing model executors. \thiswork{} therefore uses
staging buffers. For each collective, the runtime copies the local input slice
from the executor tensor into a symmetric buffer region, performs remote data
movement through the selected backend, and writes the result back to the
executor-visible output tensor.

Staging adds local copy work, but it also creates an opportunity for
pipelining. \thiswork{} divides the payload into communication chunks so that
local staging and remote data movement can be overlapped. While one chunk is
being transferred through the selected backend, the runtime can stage another
chunk between executor tensors and symmetric buffers. This hides part of the
local-copy overhead without requiring executors to allocate their tensors from a
symmetric-memory pool.

Different backends are preferable at different message sizes. CUDA-kernel
communication has low setup overhead and can combine data movement with simple
packing or reduction logic; copy-engine transfers reduce SM interference for
larger point-to-point payloads; and TMA-style transfers, when available, improve
bulk movement for structured tensor tiles. \thiswork{} uses a selector populated
from microbenchmark results to choose a backend and chunk size for
each collective pattern and message-size range.

This backend selection is hidden behind the group-free collective API. The
scheduler chooses a trajectory task's execution layout, and the executor issues
the collective required by its parallel specification. The communication runtime
then handles agreement, staging, backend selection, and data movement. As a
result, dynamic execution groups remain policy-level objects, while the
low-level communication path can adapt to hardware and message size.

\section{\thiswork{} Implementation}
\label{sec:implementation}

We implement \thiswork{} in vLLM-Omni as an event-driven control plane and a set
of execution-plane components.

\subsection{Event-Driven Control and Execution Planes}

\thiswork{} separates scheduling state from task execution. The control plane
owns request admission, trajectory task graphs, dependency state, artifact
metadata, resource availability, and policy invocation. The execution plane
contains worker processes, model executors, and communication runtimes; workers
execute assigned trajectory tasks and report events back to the control plane.

The key control-path event is dispatch completion. When the control plane
dispatches a task, the runtime resolves its logical execution group, prepares
required inputs, allocates output artifact handles, records their expected
layouts, and sends the task to participating workers. Once this CPU-side
dispatch finishes, the control plane can prepare successor state, plan later
migrations, and issue independent work while the GPU task is still running.

Device execution is tracked separately. Workers record CUDA events around task
execution, and a monitor thread reports execution-start and
execution-completion events to the control plane. These events mark device
completion, materialize output
artifacts, release worker resources, detect failures, and calibrate the runtime
cost model with measured task durations. Consumers remain blocked until their
inputs are materialized, but preparation of scheduling state does not wait for
device completion. Separating dispatch completion from device completion keeps scheduling
asynchronous with GPU execution. Policy evaluation, dependency updates,
migration planning, and other CPU-side work overlap with device execution
instead of extending the GPU critical path at every trajectory boundary.

\subsection{Model Adapters}

\thiswork{} keeps model-specific logic behind a narrow adapter interface. A
model adapter translates a diffusion pipeline into three runtime components: a
request converter, task executors, and artifact codecs. The converter maps an
incoming request to trajectory tasks and logical artifacts; executors run
assigned tasks under a chosen layout; and codecs describe logical artifact layouts
under different parallel settings.

The request converter determines the task granularity exposed to scheduling. It
emits stage tasks for pipeline components such as text encoding and VAE
decoding, as well as one task for each denoising step. Each executor accepts a
task description and execution layout, invokes the corresponding model stage,
and uses the communication runtime when its parallel specification requires
collectives.

Artifact codecs connect model-specific tensor structures to the
placement-agnostic task graph. For each logical artifact, a codec identifies
replicated, sharded, and metadata-only fields and reports the global shape and
per-rank slice of tensor fields under a given layout. The runtime uses these
views to determine whether an artifact can be consumed directly or requires
migration. This interface keeps scheduling policies independent of model internals. A
policy sees ready tasks, request metadata, resource availability, and cost
estimates, but not how latents are sharded, scheduler state is serialized, or a
model stage invokes collectives. Adding a new diffusion pipeline therefore
requires implementing an adapter rather than rewriting policies or embedding
model-specific migration logic in the scheduler.

\subsection{Layout-Aware Artifact Migration}
\label{sec:migration}

When adjacent trajectory tasks use different execution layouts, the runtime
must reconstruct their logical artifacts without exposing migration decisions
to the scheduling policy. \thiswork{} derives the required transfers from the
producer and consumer artifact views in three steps: layout exchange, migration
planning, and distributed execution.

First, the leader of the source group obtains source and destination views of
each migrated artifact from the participating ranks. The adapter's codec reports
whether each field is replicated, sharded, or metadata-only; for tensor fields,
it also reports the global shape and per-rank slice under each layout.

Second, the leader of the source group derives a migration plan from the two
views. For a sharded tensor field, the planner intersects each source-owned slice with
each destination-required slice. Every non-empty intersection becomes a
transfer entry containing the source rank, destination rank, source tensor
range, destination tensor range, and byte size.

Third, the leader of the source group distributes the plan to participating
ranks. Each rank extracts its local actions, packs required tensor ranges, exchanges
data with peers through point-to-point transfers, and installs received ranges
into the destination artifact representation. The consumer can then read the
artifact under its assigned layout.

\thiswork{} executes migration transfers through the same group-free
communication substrate, avoiding PyTorch~\cite{pytorch} point-to-point paths
that may silently construct two-rank communication groups in the background.
For each migration edge, the runtime uses a logical pair group and the
agreement mechanism from \cref{sec:gfc}; larger migrations are decomposed into
multiple transfer entries.

\subsection{Scheduling Policies}
\label{sec:policies}

\thiswork{} exposes scheduling as a policy-layer decision. A policy observes
ready trajectory tasks, request metadata, resource availability, and cost
estimates, then returns a task ordering and execution layout for each dispatched
task. We implement three representative policies.

\textbf{FCFS with workload-aware group assignment.}
This policy partitions the cluster into worker groups and serves requests in FCFS order. When a task becomes ready, it is assigned to the
feasible group with the lowest estimated queued workload. This approximates a
throughput-oriented configuration: simple request ordering with runtime-managed
load balancing across independent groups.

\textbf{SRTF with per-rank local queues.}
This policy first assigns each request to a feasible rank using estimated queued
workload. Each rank then schedules its own ready trajectory tasks by shortest
remaining trajectory work~\cite{srtf}, estimated from request shape, trajectory
position, and profiled task costs. Compared with FCFS, SRTF uses diffusion's
predictable costs to reduce mean latency, while
keeping the ordering decision local to the rank that owns the request.

\textbf{EDF with best-fit parallelism.}
This policy targets SLO settings. It orders ready tasks by earliest deadline
first~\cite{edf}, evaluates candidate layouts with the cost model, and selects
the smallest parallel configuration predicted to meet the deadline. If a
request is at risk of missing its deadline, the policy can assign a larger
group at the next trajectory boundary, demonstrating SLO-aware dynamic
placement without hard-coding deadline logic into model executors.

These policies are deliberately simple: they differ only in task ranking and
layout choice. Dependency tracking, asynchronous dispatch, dynamic-group
communication, and artifact migration are handled by the runtime.

\subsection{Simulator and Cost Model}

The same abstraction supports offline policy exploration. As shown in
\cref{sec:motivation}, the preferred scheduling policy depends on the workload
mix, request shapes, serving objective, and available ranks. A deployment may
therefore need to compare several policies before choosing one for the online
serving path.

As described in \cref{sec:predictable-execution-structure}, diffusion requests
expose predictable trajectory task graphs and layout-specific costs. The
simulator replaces worker execution with cost-model completion events while
preserving task readiness, dependency updates, resource allocation, and policy
invocation. Its cost model is populated from profiled task latencies indexed by
model, task type, request shape, and parallel configuration.

Because the simulator and online runtime share the same policy interface, a
policy selected offline can be deployed without rewriting its decision logic.
In simulation, dispatch decisions produce cost-model completion events; in the
online runtime, the same decisions are executed by GPU workers and reported as
real completion events. The simulator is therefore an alternative execution
backend for the same trajectory abstraction rather than a separate scheduling
model.

\section{Evaluation}
\label{sec:evaluation}

We evaluate \thiswork{} on image and video diffusion serving workloads.

\begin{figure*}[t]
  \centering
  \includegraphics[width=\textwidth]{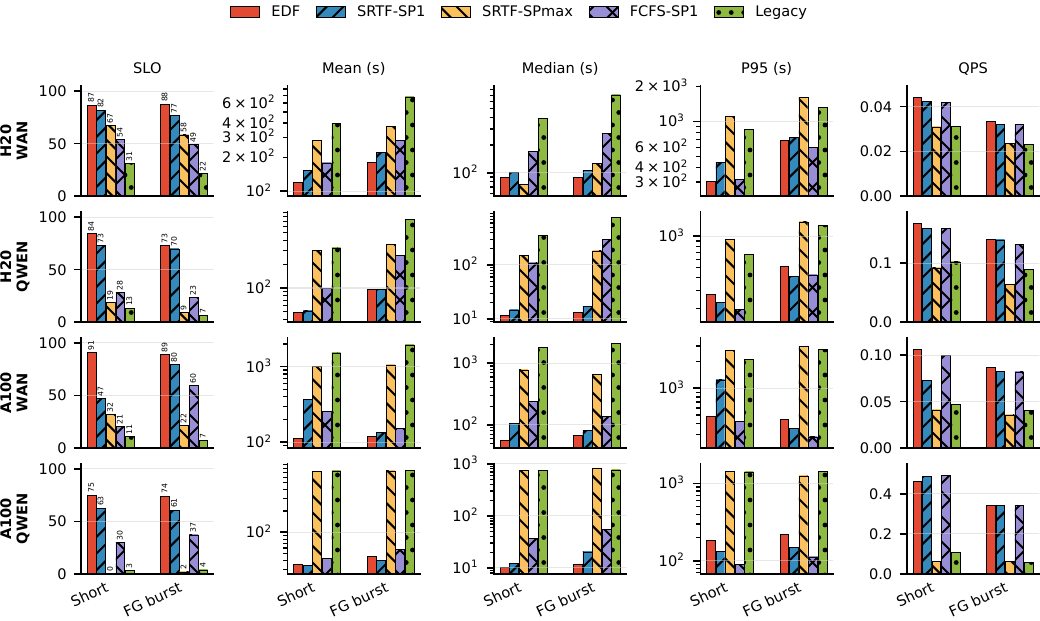}
  \caption{End-to-end serving results on H20 and A100. Each row is a
  platform-model pair, and each column reports one metric across the short and
  foreground-burst workloads. Legacy is the native vLLM-Omni fixed-pipeline
  execution path with static parallelism; the other policies are implemented on
  top of \thiswork{}. SLO
  attainment includes failed requests as violations, while latency statistics
  are computed over completed requests only.}
  \vspace{-1em}
  \label{fig:main-results}
\end{figure*}

\subsection{Experimental Setup}

We implement \thiswork{} on vLLM-Omni 0.17.0~\cite{vllm-omni} using Python
3.12, CUDA 12.9, PyTorch 2.10~\cite{pytorch}, Triton 3.6.0~\cite{triton}, and
NCCL 2.27.5~\cite{nccl}. Experiments run on a 4-GPU H20 server and an 8-GPU
A100 server, using Wan2.2-5B~\cite{wan} for video generation and
Qwen-Image~\cite{qwen-image} for image generation.

\begin{figure}
  \centering
  \includegraphics[width=\linewidth]{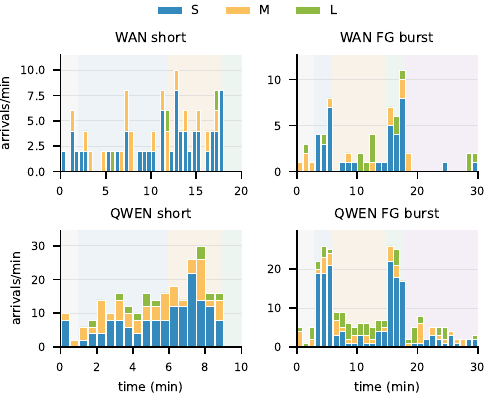}
  \caption{Arrival-rate timelines of the short and foreground-burst workloads
  used in the main results. Colors denote short (S), medium (M), and long (L)
  request classes.}
  \label{fig:workload-traces}
\end{figure}

For each model and platform, we generate fixed-duration request traces with
arrival rates calibrated from measured platform throughput, so policies are
compared under comparable serving pressure rather than a single absolute request
rate. \cref{fig:workload-traces} shows two workloads: a compact
mixed-arrival period and a foreground-burst setting where bursts of short
requests arrive while longer requests may already be in flight. Together, they
test whether the runtime can preserve mixed-request concurrency while giving
urgent short requests priority in scheduling.
For Wan2.2, short, medium, and long (S/M/L) requests generate
480$\times$832 videos with 49 frames, 480$\times$832 videos with 81 frames, and
720$\times$1280 videos with 81 frames, respectively. For Qwen-Image, S/M/L
requests generate 512$\times$512, 1024$\times$1024, and 1536$\times$1536
images.

Each request's SLO is defined relative to its profiled standalone service time
on the same model and platform. A class-$c$ request arriving at time $a$ receives
a scheduler-visible deadline $a + \alpha_c T_c$, where $T_c$ is the profiled
service time of its request shape. We use
$\alpha_{\mathrm{S/M/L}}=2.0/2.5/3.5$ for Wan2.2 and $1.5/2.0/6.0$ for
Qwen-Image, giving short interactive requests tighter relative deadlines while
leaving longer requests more queueing slack. The multipliers are fixed across
platforms and workloads, while $T_c$ is re-profiled per platform. We choose
these representative multipliers before comparing policies and do not tune
them per policy. The client-visible deadline additionally includes a fixed
allowance for non-profiled overheads: 5~s for Wan2.2 and 1~s for Qwen-Image. We
report SLO attainment as the fraction of submitted requests that complete by
this deadline.

Each request also has a loose client-side timeout, 1500~s for image generation
and 3600~s for video generation. Timed-out requests are recorded as failures and
SLO violations. Latency statistics include only completed requests, and
throughput is measured as completed requests per second, so latency results for
policies with failures are optimistic.

\subsection{Baselines}

We compare against the native vLLM-Omni serving path, denoted Legacy, and the
policies from \cref{sec:policies}. Legacy uses vLLM-Omni's original
fixed-pipeline execution with static parallelism over the full machine. EDF,
SRTF-SP1, SRTF-SPmax, and FCFS-SP1 are implemented on top of the \thiswork{}
policy interface.

\subsection{Main Results}

\cref{fig:main-results} shows that different objectives favor different
policies. EDF achieves the highest SLO attainment in every
platform-model-workload setting, with large gains in bursty or highly
concurrent cases: on A100 Wan2.2 foreground burst, SLO attainment improves from
7.5\% with Legacy to 88.8\% with EDF; on H20 Qwen-Image foreground burst, it
improves from 6.2\% to 72.8\%. Across all main-result settings, \thiswork{}
improves throughput by up to 6.01$\times$, reduces mean latency by up to
95.3\%, and reduces SLO violation rate by up to 89.6\% over Legacy. EDF improves
deadline attainment because it orders trajectory tasks by deadline and chooses
parallelism from estimated completion time instead of using one fixed execution
order and static parallel configuration.

The latency metrics explain why no single policy dominates all objectives.
Legacy and SRTF-SPmax can reduce the service time of individual large tasks
with large execution groups, but they reduce concurrency and increase queueing under mixed
workloads. Smaller single-rank policies can therefore win on latency:
SRTF-SP1 gives the lowest mean latency on several Qwen-Image settings, while
FCFS-SP1 gives the lowest P95 latency on the A100 Qwen-Image short workload.
These results support the main design goal of \thiswork{}: scheduling strategy
should be a policy-layer choice over one runtime substrate.

The loss of concurrency also causes a capacity collapse for the large-group
policies on A100 Qwen-Image. On the foreground-burst workload, Legacy and
SRTF-SPmax complete only 37\% and 39\% of submitted requests, respectively,
while the other policies complete nearly all requests. These failures occur
when requests remain in the serialized SP8 queue beyond the client timeout.
Because failed requests are excluded from latency statistics, the reported
latencies for these baselines are optimistic.

\begin{figure}
  \centering
  \includegraphics[width=\linewidth]{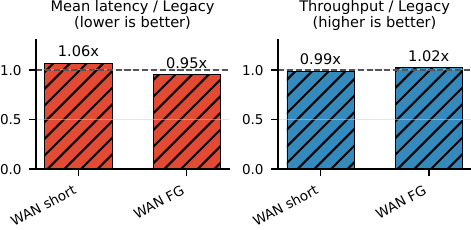}
  \caption{Runtime overhead relative to the native Legacy path on 4-GPU H20.
  FCFS-SP4 pins \thiswork{} to the same FIFO order and static SP4 layout.
  }
  \vspace{-1em}
  \label{fig:runtime-overhead}
\end{figure}

\subsection{Runtime Overhead}

\cref{fig:runtime-overhead} compares the native Legacy path, which uses
fixed-pipeline execution with static parallelism, with \thiswork{} pinned to
FCFS-SP4, which uses the same FIFO order and a
full-machine SP4 group. FCFS-SP4 closely matches Legacy in throughput and mean
latency, showing that policy programmability introduces negligible overhead
when the execution order and layout are fixed.

\begin{figure}
  \centering
  \includegraphics[width=\linewidth]{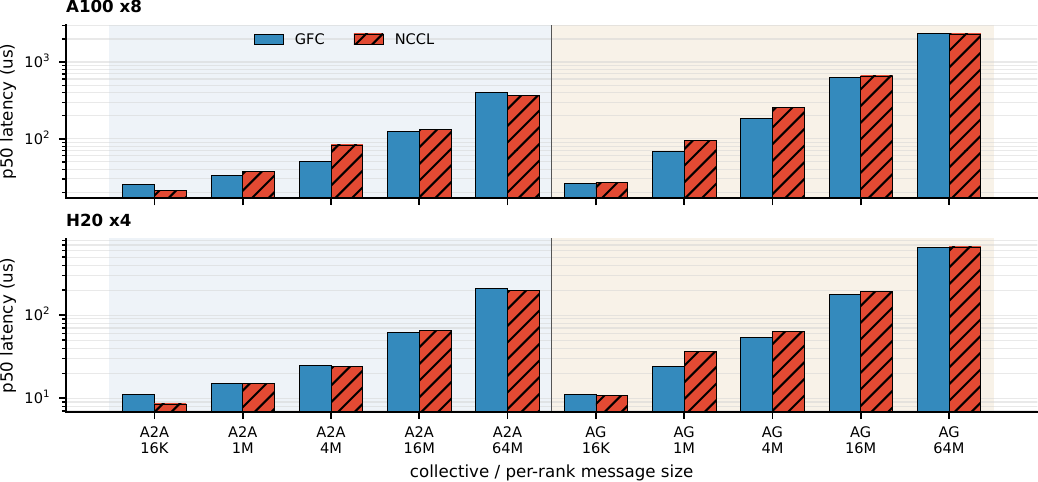}
  \caption{Latency of GFC and NCCL collectives on A100 and H20.
  We measure BF16 all-to-all (A2A) and all-gather (AG) across different
  per-rank message sizes.}
  \label{fig:gfc-latency}
  \vspace{-1em}
\end{figure}

\subsection{Group-Free Collective Performance}

\cref{fig:gfc-latency} compares steady-state GFC latency against NCCL with
pre-initialized process groups. For small messages, GFC can be slower because
its fixed synchronization and staging overheads are not yet optimized for this
regime. Diffusion-serving collectives typically use larger intermediate tensors,
for which these costs are amortized; in this range, GFC generally matches or
outperforms NCCL. For example, GFC reduces A100 4~MB all-to-all latency from
83.0~$\mu$s to 50.4~$\mu$s and H20 1~MB all-gather latency from 36.5~$\mu$s to
24.1~$\mu$s. Thus, GFC avoids serving-path process-group construction while
maintaining competitive steady-state performance.

\begin{figure*}
  \centering
  \includegraphics[width=\textwidth]{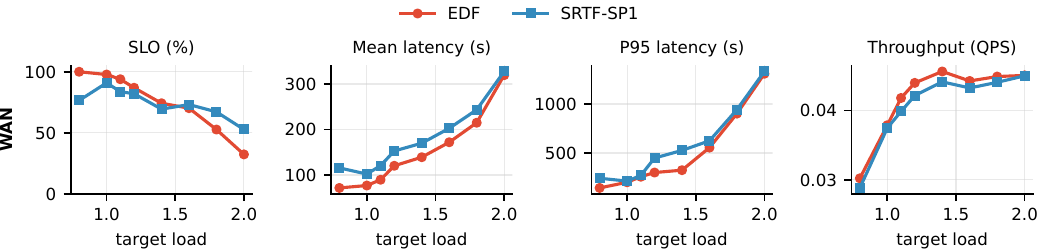}
  \vspace{-1em}
  \caption{Performance of EDF and SRTF-SP1 under increasing arrival rates on
  the 4-GPU H20 Wan2.2 short workload. Target load is normalized to the
  estimated serving capacity.}
  \vspace{-1em}
  \label{fig:arrival-scaling}
\end{figure*}

\subsection{Scaling with Arrival Rate}

\cref{fig:arrival-scaling} evaluates EDF and SRTF-SP1 as arrival rate
increases. At low and moderate loads, EDF achieves higher SLO attainment by
prioritizing earlier deadlines and adding parallelism when needed. Under
sustained overload, however, EDF may keep rescuing tight-deadline requests with
larger sequence-parallel groups, consuming more GPUs per request and delaying
new arrivals. SRTF-SP1 then achieves higher SLO attainment by preserving
single-rank concurrency and completing more requests with little remaining
work. The crossover shows that deadline-aware parallelism helps when capacity is
available but can reduce overall SLO attainment under overload.

\subsection{Simulator Fidelity}

\begin{figure}[t]
  \centering
  \includegraphics[width=\linewidth]{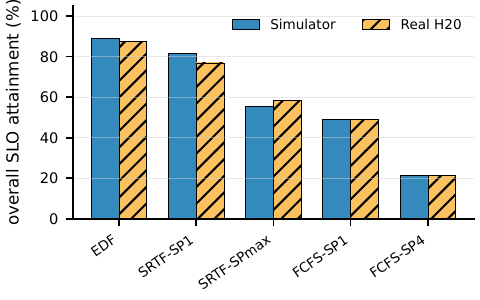}
  \vspace{-1.5em}
  \caption{Simulator versus real 4-GPU H20 overall SLO attainment for the
  Wan2.2 foreground-burst workload. The simulator replays the exact request
  trace and policy logic using measured stage costs.}
  \label{fig:sim-vs-real}
  \vspace{-1em}
\end{figure}

\cref{fig:sim-vs-real} compares simulator predictions with real serving runs on
a representative H20 workload. We compare the four programmable policies from
the main results and include FCFS-SP4 as a static large-group baseline.
Simulator predictions closely match real serving results: all five policies
differ by at most 4.7 percentage points in overall SLO attainment. The
simulator is therefore accurate enough to compare policy trends and identify
candidates, while final selection still requires real-system validation.

\section{Related Work}

\paragraph{Diffusion model serving.}
Recent diffusion serving systems exploit stage- or step-level execution structure to improve serving efficiency. TridentServe dynamically allocates resources across encoding, denoising, and decoding stages~\cite{TridentServe}, while TetriServe and GenServe perform step-level scheduling for homogeneous and heterogeneous diffusion workloads~\cite{TetriServe,GENSERVE}. Other systems focus on batching, mixed-resolution serving, model cascades, or adapter-intensive workloads~\cite{ditserve,PATCHEDSERVE,MixFusion,diffserve,katz}. In contrast, \thiswork{} introduces a programmable elastic runtime that treats GPU parallelism as a schedulable resource and supports dynamic parallelism adaptation under diverse serving objectives. Existing DiT inference systems provide efficient parallel execution layouts, including sequence, patch-pipeline, and CFG parallelism~\cite{xdit,pipefusion,CompactFusion}. Frameworks such as vLLM-Omni, SGLang Diffusion, and FastVideo integrate these techniques into practical deployment stacks~\cite{vllm-omni,sglang-diffusion,fastvideo}. \thiswork{} is complementary to these efforts: rather than proposing new parallel kernels, it enables existing execution layouts to be dynamically selected and reconfigured at runtime.

% LLM serving systems leverage token-level scheduling opportunities enabled by autoregressive decoding, supporting batching, preemption, migration, and elastic execution~\cite{orca,vllm,sglang,sarathi,distserve,mooncake,fastserve,llumnix,loongserve,nanoflow,parrot,preble}. In contrast, DiT workloads lack token-prefix semantics because each denoising step performs bidirectional computation over the latent sequence. \thiswork{} therefore introduces execution-unit-based scheduling and runtime-managed parallelism tailored to diffusion workloads.

\paragraph{Collective communication and dynamic groups.}
Collective communication libraries such as NCCL, RCCL, UCC, and UCCL rely on statically defined communication groups~\cite{nccl,rccl,ucc,uccl,ncclx}, while recent work improves collective performance through synthesis, scheduling, and topology-aware optimization~\cite{msccl,sccl,taccl,tacos,dfccl,deepep,uccl-ep}. These systems optimize communication within a predefined group. In contrast, \thiswork{} addresses the serving-runtime challenge of online execution-group reconfiguration through group-free collectives, eliminating expensive communicator construction on the serving path.

\section{Conclusion}

This paper presented \textbf{\thiswork{}}, a policy-programmable runtime for elastic DiT serving. By treating GPU parallelism as a first-class schedulable resource, \thiswork{} enables online adaptation of resource allocation to workload demands, system conditions, and service objectives. Through an asynchronous execution abstraction, group-free collectives, and a programmable runtime substrate, \thiswork{} makes elastic parallelism practical for real-world DiT serving workloads, substantially improving throughput, latency, and SLO attainment over fixed-pipeline execution with static parallelism. More broadly, we believe that future generative AI serving systems should move beyond static resource allocation and embrace runtime-managed parallelism as a fundamental systems abstraction.

% bibliography
\bibliographystyle{ACM-Reference-Format}
\bibliography{ref.bib}

\end{document}